# Managing Query Compilation Memory Consumption to Improve DBMS Throughput

Boris Baryshnikov, Cipri Clinciu, Conor Cunningham, Leo Giakoumakis, Slava Oks, Stefano Stefani

Microsoft Corporation
One Microsoft Way
Redmond, WA 98052 USA
{borisb,ciprianc,conorc,leogia,slavao,stefanis}@microsoft.com

## Abstract

While there are known performance trade-offs between database page buffer pool and query execution memory allocation policies, little has been written on the impact of query compilation memory use on overall throughput of the database management system (DBMS). We present a new aspect of the query optimization problem and offer a solution implemented in Microsoft SQL Server 2005. The solution provides stable throughput for a range of workloads even when memory requests outstrip the ability of the hardware to service those requests.

## 1. Introduction

Memory Management is a critical component of DBMS performance. In modern systems, memory trade-offs are far more complex than the classic problems of database page buffer management or reserving memory for hashes and sorts during query execution [5]. Current systems use more ad-hoc queries that make query compilation memory more important in memory reservation policies. Furthermore, these ad-hoc deployments make it harder to provision hardware, and thus they are more often run at or beyond the capabilities of the underlying hardware. This requires intelligent trade-offs among other memory consumers in a DBMS, as every byte consumed in query compilation, query plan caches, or other components effectively reduces the available memory for caching data pages or executing queries.

In our research we identified compile-intensive ad-hoc workloads that consume enough memory in query compilation to disturb the traditional memory consumption trade-offs between a database page buffer pool and query execution. For these scenarios, overall system throughput was significantly reduced due to memory thrashing among components. Even if the system has enough memory to service multiple simultaneous query compilations, allowing all of them to occur at the same time might not maximize throughput. Excessive concurrent compilation memory usage steals a significant number of pages from the database page buffer pool and causes increased physical I/O, reduces memory for efficient query execution, and causes excessive eviction of compiled plans from the plan cache (forcing additional compilation CPU load in the future). The interplay of these memory consumers is complex and has not been adequately studied.

In this paper we present a solution to the above problem by providing a robust query compilation planning mechanism that handles diverse classes of workloads, prevents memory starvation due to many concurrent query compilations, and dynamically adjusts to load to make intelligent memory trade-offs for multiple DBMS subcomponents that improve overall system reliability and throughput. This solution has been implemented and tested against Microsoft SQL Server 2005 and is part of the final product.

## 2. Memory Influence on Performance

As many DBMS installations run on dedicated hardware, the rest of this paper will assume, for simplicity, that almost all physical memory is available to the DBMS and that it is the only significant memory-consuming process being run on the hardware.

### 2.1 DBMS Subcomponent Memory Use

DBMS subcomponent contention over memory can impact system throughput, and managing the interplay of these components to improve system performance can be challenging. Each DBMS subcomponent uses memory differently, and this can impact the heuristics and policies required to maximize overall performance.

Query compilation (and, more specifically, query optimization) consumes memory differently than other DBMS subcomponents. Many modern optimizers consider a number of functionally equivalent alternatives and choose the final plan based on an estimated cost function. This entire process uses memory to store the different alternatives for the duration of the optimization process. The memory consumed during optimization is closely related to the number of considered alternatives. For an arbitrary query, the total memory consumed during optimization is often hard to predict due to the large number of alternatives and various optimization algorithms employed. This makes it difficult to understand memory consumption performance trade-offs in relation to other DBMS subcomponents. While work has been done on dynamic query optimization, where the number of considered alternatives (and thus the amount of memory consumed) is related to the estimated cost function for the query, no work, to our knowledge, has been done on the value of consuming more memory optimizing a query in a memory-constrained workload.

## 2.2 DBMS Design vs. System Throughput

The combination of limited memory/virtual address space and a diverse set of memory consumers poses challenges to the DBMS implementer when trying to guarantee good system throughput. A naïve approach of placing caps on each memory subcomponent to avoid memory starvation does not always work due to the varied nature of workloads and the inability of the system to plan for work in the future without additional guidance from the user. Even within a subcomponent, memory allocation policies can be difficult to tune to achieve system stability. For example, if many large queries are compiling simultaneously, each compilation can consume a significant fraction of system memory. Query compilations can deadlock on each other if both are waiting for memory consumed by another compilation. Even if the system aborts most of these queries to allow a few to complete, those aborted queries likely need to be resubmitted to the system.

In absence of a central controlling mechanism, the overall system will likely either not perform well or not be stable in all situations. Making proper decisions to receive, evaluate, and arbitrate requests for memory amongst multiple consumers can improve system stability and increase throughput, even when the system is running at or beyond the capabilities of the hardware. This approach is described in more detail in the following section.

## 3. Memory Broker

We propose a "Memory Broker" to manage the physical memory allocated to DBMS subcomponents. The broker accounts for the memory allocated by each subcomponent, recognizes trends in allocation patterns, and provides the mechanisms to enforce policies for resolving contention both within and among subcomponents. This subcomponent enables a DBMS to make better global decisions about how to manage memory and ultimately achieve improved throughput.

The Memory Broker monitors the total memory usage of each subcomponent and predicts future memory usage by identifying trends. If the system is not using all available physical memory, no action is taken and the system behaves as if the Memory Broker was not there. If the future memory total is expected to exceed the available physical memory, the broker predicts the actual amount of physical memory that the subcomponent should be able to allocate (accounting for requests from other subcomponents). The broker also sends notifications to each subcomponent with its predicted and target memory numbers and informs that subcomponent whether it can continue to consume memory, whether it can safely allocate at its current rate, or whether it needs to release memory. In our implementation, the overhead of this mechanism is extremely small. It is still possible to have out-of-memory errors if many subcomponents attempt to grow simultaneously. The system relies on the ability of various subcomponents to make intelligent decisions about the value of optional memory allocations, free unneeded or low-value memory, and reduce the rate of memory allocations over time.

The Memory broker provides an indirect communication channel for one subcomponent to learn about the overall memory pressure on the system. It also helps to mitigate "wild" swings in subcomponent memory allocations and tends to make the overall DBMS behave more reliably by reducing probability of aborting long-running operations such as compiling and/or executing a query.

DBMS subcomponents impose different requirements on a memory subsystem through their usage patterns that can impact how the Memory Broker operates. For example, the database page buffer pool contains data pages that have been loaded from disk. Replacement policies can be used to remove older pages to load currently needed pages, but they can also be used to enable the buffer pool to identify candidates necessary to shrink its size. Other caches can support shrinking using the same technique. The memory consumed during query execution is usually predictable as many of the largest allocations can be made using early, high-level decisions at the start of the execution of a query. Unlike caches, however, the execution of queries may require that memory be allocated for the duration of the query. Therefore, the subcomponent may be less capable to respond to memory pressure from a Memory Broker at any specific time. However, it can potentially respond to memory pressure based on the shape of the query and the relative position of the operators being executed.

Query compilation also uses memory in ways interesting to a Memory Broker component, and this is discussed in detail in the next section.

## 4. Query Compilation Throttling

Query compilation consumes memory as a function of both the size of the query tree structure and number of alternatives considered. Beyond dynamic optimization, which has traditionally been based on estimated query runtime and not memory needs, there are no published techniques to avoid memory use during query compilation for standard approaches.

Our analysis of actual compile intensive workloads showed that high memory consumption is typically caused by several medium/large concurrent ad hoc compilations, rather than one or few very large queries (that are less likely to be executed in a highly concurrent environment where different connections and components compete for memory). While it may not be easy (or desirable) to modify the main optimization process to account for memory pressure, it is possible to change the *rate* at which concurrent optimizations proceed to respond to memory pressure. In this section, we describe a query compilation planning mechanism that handles multiple classes of workload goals, dynamically adjusts to system memory pressure, and interacts with the dynamic programming algorithms used in many modern optimizers to make intelligent decisions about memory use during the compilation process. This system improves overall system throughput and reduces resource errors returned to clients when the system is under memory pressure.

### 4.1 Solution Overview

We propose a query compilation throttling solution that responds to memory pressure by changing the rate at which compilations proceed. If we assume that memory use roughly grows with compilation time, throttling at least some compilations restricts the overall memory usage by the query optimization subcomponent and can improve the system throughput. Blocked compilations wait for resources to become available before continuing. If the compilation of a query remains blocked for an excessively long period of time, its transaction is aborted with a "timeout" error returned to the client. Properly tuned, this approach allows the DBMS implementer to achieve a balance between out-of-memory errors and throttle-induced timeouts for a variety of workloads. Our approach gives preference to compilations that have made the most progress and avoids many cases where a compilation is aborted after completing most, but not all, of the compilation process.

Blocking is implemented through a series of monitors that are acquired during the compilation. The blocking is tied to the amount of memory allocated by the task instead of specific points during the query compilation process. This provides a more robust mechanism to control the impact of compilation on overall system memory load over a wide variety of schema designs and workload categories. These monitors contain progressively higher memory thresholds and progressively lower limits on the number of allowed concurrent compilations as illustrated in Figure 1. The monitors are acquired sequentially by a compilation as memory usage for that task increases and are released in reverse order. If memory is not available at the time of acquisition, the compilation process is blocked until memory becomes available when other compilations, executions, or memory requests elsewhere in the system are completed. A timeout mechanism is used (with increasing timeouts for later monitors) to return an error to the user if the system is so overloaded that a compilation does not make any progress for a long period of time.

**Figure 1 Memory Monitors**

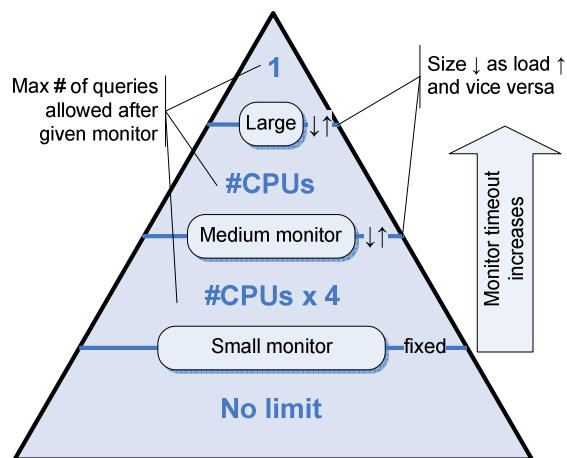

Restraining compilations effectively avoids some cases where many simultaneous compilations consume a disproportionately high fraction of the available physical memory. Since memory is a scarce resource, preserving some fraction of it for use by the database page buffer pool and query execution allows these components to more efficiently perform their functions. Blocking some queries can reduce the need for other subcomponents to return memory from caches if many large, concurrent compilations occur. This can spread memory use over time instead of requiring other subcomponents to release memory. The intended goals of this approach are to improve maximum throughput and to enable that throughput to work for larger client loads on the system.

Our implementation uses three monitors. Experimental analysis showed that dividing query compilations into four memory usage categories gives the best balance between trying to handle different classes of workloads and limiting the compilation time overhead of the mechanism. Query compilations that consume less memory than the first monitor threshold proceed unblocked. The first threshold is configured differently for each supported platform architecture to allow a series of small diagnostic queries to proceed without acquisition of the first (smallest) monitor. This enables an administrator to run diagnostic queries even if the system is overloaded with queries consuming every available 'slot' in the memory monitors. The first monitor allows four concurrent compilations per CPU and is used to govern "small" queries. Typically, most OLTP-class queries would fall into this category. The second monitor is required for larger queries, allowing one per CPU. Many TPC-H queries, which require the consideration of many alternatives, would be in this category. The final governs the largest queries and allows only one at a time to proceed. This class of query uses a sizable fraction of total available memory on the system. The largest memory-consuming queries are serialized to avoid starvation of other subcomponents and allow the query to complete compilation. This approach allows us to restrict compilation, in most cases, to a reasonable faction of total memory and allow other subcomponents to acquire memory.

**Figure 2 Compilation Throttling Example**

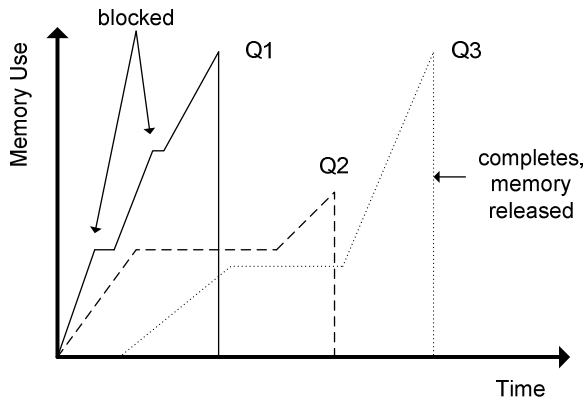

Figure 2 contains a simplified example to describe how query compilation throttling might work in practice. In this example, Q1 and Q2 start compiling at approximately the same time. However, Q1 consumes memory at a faster rate than Q2. This could occur if the query was larger, the tables contained more complex schemas, or perhaps that thread of control received more time to execute. Q1 then blocks at the first monitor, as denoted by the first flat portion of the graph of Q1's memory use. This occurred because other queries (not shown in the example) were consuming enough resources to induce throttling. Once enough memory is available, Q1 continues executing, blocks again at the next monitor, eventually is allowed to continue, and finally finishes compilation. At the end of compilation, memory used in the process is freed and the query plan is ready for execution. Q2 executes in a similar manner. It waits much longer to acquire the first monitor (meaning that the system is under more memory pressure or that other compilations are concurrently executing and using memory). Q2 finishes and frees memory, but it did not require as much memory as Q1 to compile. In this example, Q3 is actually blocked by Q2 and only proceeds once Q2 is finished and releases its resources. From the perspective of the subcomponent author, the only perceptible difference in this process from a traditional, unblocked system is that the thread sometimes receives less time for its work. The individual thread scheduling choices are made by the system based on global goals that effectively prioritize earlier over later compiles when making scheduling (and thus allocation) decisions.

We have also extended this approach with two novel extensions. First, we have made the monitor memory thresholds for the larger gateways dynamic. This is based on the broker memory target. This allows the system to throttle some workloads more aggressively when other subcomponents are heavily using memory, making the system even more responsive to memory pressure. The thresholds are computed attempting to divide the overall query compilation target memory across the categories identified by the monitors. For example, the second monitor threshold is computed as [target] * F / S, where F and S are respectively the fraction of the target allotted to and the current number of small query compilations. In other words, small queries together can consume up the F fraction of the target, after which the top memory consumers are forced to upgrade to the medium category. The values of the F fractions were identified with a long process of tuning and experimentation against several actual workloads. We also have leveraged the notification mechanisms to determine that the system will likely run out of memory before compilation completes. When this happens, we can return the best plan from the set of already explored plans instead of simply returning out-of-memory errors. Both techniques allow the system to better handle low-memory conditions.

## 5. Experimental Results

Standard database benchmarks (TPC-H, TPC-C [6]) contain queries with moderate or small memory requirements to compile. Large decision support systems run queries with much higher resource requirements. To evaluate our solution, we developed a benchmark based on a product sales analysis application created by a SQL Server 2005 customer. For the purposes of this paper, we will refer to that benchmark as the SALES benchmark.

## 5.1 SALES Benchmark

The SALES application is a Decision Support System (DSS) which uses a large data warehouse to store data from product sales across the world. This application submits almost exclusively ad-hoc queries over significant fractions of the data. Many users can submit queries simultaneously. The customer runs a number of large-CPU systems at or near capacity to handle their user query load due to their unpredictable, ad-hoc workload.

The SALES benchmark uses a somewhat typical data warehouse schema, meaning that it has a large fact table and a number of smaller dimension tables. The largest fact table from the database contains over 400 million rows. The "average" query in this benchmark contains between 15 and 20 joins and computes aggregate(s) on the join results. As a comparison, TPC-H queries contain between 0 and 8 joins with similar numbers of indexes per table. The data mart in our experiments contains a snapshot of the data from the customer's application and is 524 GB in size.

We executed this benchmark against SQL Server 2005. It features dynamic optimization, meaning that the time spent optimizing a query is a function of the estimated cost of the query. Therefore, more expensive queries receive more optimization time. In our experiments, the queries in the SALES benchmark use one to two orders of magnitude more memory than TPC-H queries of similar scale.

Our benchmark models the basic functionality of the application and contains 10 complex queries that are representative of the workload. To simulate the large number of unique query compilations, our load generator modifies each base query before it is submitted to the database server to make it appear unique [7] and to defeat plan-caching features in the DBMS.

## 5.2 Execution Environment/Results

We execute the SALES benchmark using a custom load generator which simulates a number of concurrent database users who submit queries to the database server. For these experiments, we use a server with 8 Intel Xeon (32-bit) 700 MHz x86-based processors and 4GB of main memory. The server is using 8 SCSI-II 72GB disks configured in as a single RAID-0 drive array on a 2-channel, 160 MB/channel Ultra3/Ultra2 SCSI controller. The software on the machine is Microsoft Windows 2003 Enterprise Edition SP 1 and Microsoft SQL Server 2005. This system is a typical medium-large server installation and should reasonably reflect the hardware on which scaling problems are currently being seen in DBMS installations today.

Queries in this benchmark generally compile for 10-90 seconds and execute for 30 seconds to 10 minutes. In each subcomponent, these queries consume nontrivial amounts of memory to compile and execute. They also access large fractions of the database and thus put pressure on the database page buffer pool. Therefore, these subcomponents are actively competing for memory during the execution of this benchmark. The probability that a query will be aborted due to memory shortages is high, and the cost of each failure is also high (as the work will be retried). This places a high value on biasing resource use towards those operations likely to succeed on the first attempt.

Our experiments measure the throughput and reliability of the DBMS while running both *at and beyond* the capabilities of the hardware. "Throughput" in this context means the number of queries successfully completed per unit of time. Through experimentation, we determined that this benchmark produces maximum throughput with 30 clients on this hardware configuration. Throughput is reduced with fewer users. Increasing the number of users beyond 30 saturates the server and causes some operations to fail due to resource limitations. To measure the effect of running the system under memory pressure, we performed experiments using 30, 35, and 40 clients.

The benchmark imposes extreme loads on the server, and it takes some time for the various structures in each subcomponent to warm up and become stable enough to measure results. The results presented in this section do not include this warm-up period and the data starts at an intermediate time index. There is some fluctuation in the numbers reported because of the different sizes of the queries being executed and the non-deterministic interplay of a number of different clients attempting to proceed at once in a complex system. Experiments were run multiple times, and the results were repeatable for all types of runs presented.

### 5.2.1 Throughput Results

Figure 3 presents throughput results for the query workload for 30 clients. For each graph, the darker line with diamond points represents the results when throttling was enabled. The lighter line with square points represents the non-throttled data. The points represent the number of successful query completions since the last point in time.

Throttling improves overall throughput by approximately 35% for the 30 client case, allowing a sustained completion of 30-40 queries per time slice in the benchmark. Un-throttled compilations in this benchmark will consume most available memory on the machine and starve query execution memory and the buffer pool. Throttling also helps the 35 and 40 client cases. As visible in Figure 4 and Figure 5, the throughput is lower in each of these cases when compared to the 30 client case. However, throttling still improves throughput for a given number of clients for each of these client loads.

**Figure 3 Throughput - 30 clients**

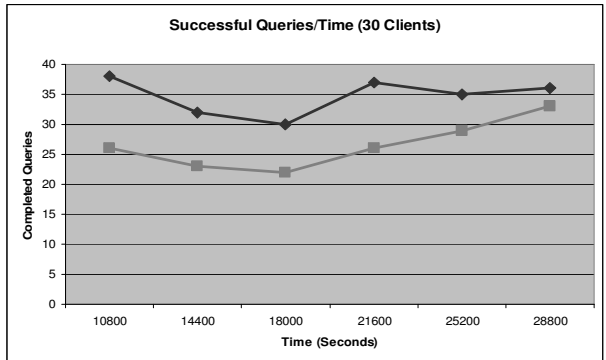

**Figure 4 Throughput - 35 clients**

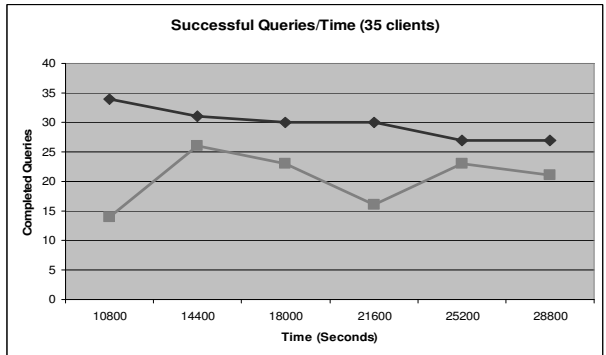

**Figure 5 Throughput - 40 clients**

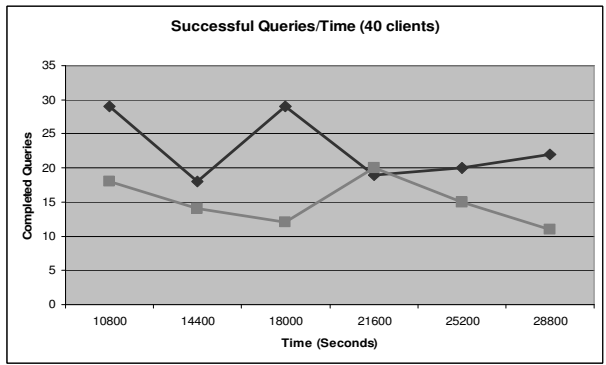

Since the data volumes are very large in this benchmark, almost every complex execution operation is performed via hashing. Therefore, each query execution is bound by the maximum size of these hash tables and the CPU work required to build and probe these structures. We also have experimental results that demonstrate measurably higher completion rates for queries when we use the throttling techniques presented in this paper.

## 6. Related Work

Much work has been done on database page buffer pool allocation/replacement strategies and execution/buffer pool trade-offs, however neither of these works specifically address compilation memory or memory from other DBMS caches. [2] and [5] are representative of the field. [5] discusses the trade-offs associated with query execution memory and buffer pool memory. [2] covers different execution classes for different kinds of queries and fairness across queries. [3] discusses the integration of cache state into query optimization. [1] covers the concept of cross-query plan fragment sharing.

## 7. Conclusions

We introduce a new form of memory/performance trade-off related to many concurrent query compilations and determine that using excessive amounts of memory in a DBMS subcomponent can impact overall system performance. By making incremental memory allocations more "expensive", we can introduce a notion of cost for each DBMS subcomponent that enables more intelligent heuristics and trade-offs to improve overall system performance. Our approach utilizes a series of monitors that restrict the future memory allocations of query compilations, effectively slowing their progress.

In our experiments, we demonstrate that throttling query compilations can improve overall system throughput by restricting compilation memory use to a smaller fraction of overall memory, even in ad-hoc workloads. This improves overall throughput and increases service reliability, even under loads beyond the capability of the hardware. In our experiments, we were able to improve system throughput by 35%.